\documentclass[11pt,twoside]{article}

\usepackage{baaa-engb2}

\usepackage{graphicx}

\usepackage[T1]{fontenc} 

\usepackage{latexsym}
\usepackage{verbatim}

\begin{document}
\def\tablename{Tabla}%

\markboth{J.C. Muzzio}%
{Chaos in elliptical galaxies}

\pagestyle{myheadings}
%
%
\vspace*{0.5cm}
\parindent 0pt{LECTURE}

\title{Chaos in elliptical galaxies}

\author{J.C. Muzzio$^{1,2}$}

\affil{%
  (1) Facultad de Ciencias Astron\'omicas y Geof\'isicas - UNLP\\
  (2) Instituto de Astrof\'isica de La Plata (CCT CONICET La Plata - UNLP)\\
}

\begin{abstract} 
Here I present a review of the work done on the presence and effects of chaos
in elliptical galaxies plus some recent results we obtained on this subject.
The fact that important fractions of the orbits that arise in potentials
adequate to represent elliptical galaxies are chaotic is nowadays undeniable.
Alternatively, it has been difficult to build selfconsistent models of elliptical
galaxies that include significant fractions of chaotic orbits and, at the same
time, are stable. That is specially true for cuspy models of elliptical galaxies
which seem to best represent real galaxies. I argue here that there is no
physical impediment to build such models and that the difficulty lies in the method
of Schwarzschild, widely used to obtain such models. Actually, I show that there
is no problem in obtaining selfconsistent models of elliptical galaxies, even cuspy
ones, that contain very high fractions of chaotic orbits and are, nevertheless,
highly stable over time intervals of the order of a Hubble time.
\end{abstract}

\section{Models of elliptical galaxies}
How can we build a dynamical model of an elliptical galaxy? The basic constituents
of galaxies are dark matter, stars and gas (plus a little dust). Gas dynamics plays
a significant role in the formation process of all galaxies, but their influence
in full-fledged elliptical galaxies is negligible, so that only dark matter and
stars need to be taken into account for models that represent them. Thus, Newtonian
dynamics is all that is needed to model elliptical galaxies and even a black hole
can be represented simply as a point mass for our purposes.

The effects of star-star interactions can be neglected for these models, as is
shown in any textbook on galactic dynamics (see, e.g., Binney \& Tremaine 2008),
and we can assume that there is a smooth distribution of mass density (dark matter
plus stars) that creates a similarly smooth potential. Let us consider now the
types of orbits we found in elliptical galaxies.

\subsection{Typical orbits in elliptical galaxies}
A significant breakthrough in elliptical galaxies dynamics was done with the use of
St\"ackel potentials that are fully integrable, so that all orbits are regular. In
particular, de Zeeuw \& Lynden-Bell (1985), de Zeeuw (1985) and Statler (1987)
investigated the "perfect ellipsoid", whose density is given by:

\medskip
\begin{equation}
\label{(Ec. perfect)}
         \rho(r) =\frac{\rho_{o}}{(1+m^{2})^{2}}  
         \,
\end{equation}
\medskip

where $m$ is the ellipsoidal radius:

\medskip
\begin{equation}
\label{(Ec. perfect)}
         {m^{2}} = \frac{x^{2}}{a^{2}} + \frac{y^{2}}{b^{2}} + \frac{z^{2}}{c^{2}}.    
         \,
\end{equation}
\medskip

Although that model is no longer regarded as an adequate representation of
elliptical galaxies, its main kinds of orbits are also found in the more
realistic models we use nowadays. They are: a) Boxes, that can come as
close as one wishes to the center of the model and do not keep the sense
of rotation; b) Minor axis tubes, that never come close to the center and
rotate always in the same sense around the minor axis of the model; c) Major
axis tubes, that never come close to the center either and always rotate in
the same sense around the major axis of the model, they are in turn divided
in inner and outer major axis tubes.

There is no reason, other than mathematical simplicity, to adopt the perfect
ellipsoid, or any other St\"ackel potential for that matter, to model
elliptical galaxies. On the contrary, it has a flat central density distribution
and we know nowadays that cuspy galaxies are the rule (we will return to this
point later on). In cuspy models the box orbits tend to be replaced by "boxlets",
i.e., resonant orbits that avoid the center of the model (Miralda-Escud\'e \&
Schwarzschild 1989).

All these are regular orbits and, of course, in generic potentials we have also
chaotic orbits but, since chaos is the main subject of the present lectures,
they will be analyzed in much more detail later on.

\subsection{The problem of self-consistency}
The galactic models must be self-consistent, that is, the mass distribution
creates a gravitational field where certain kinds of orbits are possible, and
those orbits should be such that they yield that mass distribution. The Poisson
and Boltzmann equations can be used to obtain simple spherical or disk models
(see, e.g., Binney and Tremaine 2008), but special methods are required for
the triaxial models needed for elliptical galaxies.

One of the most popular of those methods is the one due to Schwarzschild (1979).
First, one chooses a reasonable mass distribution and obtains its potential. A
library of orbits is then computed in that potential using a variety of initial
conditions in order to cover all the possible orbits in that potential, and weights
are assigned according to the time that a particle on each orbit would spent
in every region of the available space. Those weights are subsequently used
to set a system of linear equations that related the density in every region to
the fraction of orbits that make up the system. Finally, solving that system,
the fraction of every kind of orbit is obtained. 

This method certainly offers
a straightforward way of obtaining a self-consistent model, but there are
several details that should be taken into account. To begin with, there is not
a single solution, and one does not know {\it a priori} towards which solution
the method will converge. Besides, the chosen initial mass distribution conditions
the results, for example, choosing the mass distribution of the perfect ellipsoid
given above will force the system to have constant axial ratios, $b/a$ and $c/a$,
from the center through the outermost regions of the model. It is not easy to
guarantee that all the possible kinds of orbits have been included in the
library and, to make this problem worse, the libraries are usually not very large,
from several hundreds of orbits at Schwarschild's time and up to a few tens of
thousands nowadays (see, e.g., Capuzzo-Dolcetta {\it et al.} 2007).  

The most serious problem to the method of Schwarzschild is the one posed by chaotic
orbits, which can display very different behavior at different time intervals,
even over long periods of time. Sticky orbits, in particular, can mimic regular
orbits over long intervals, only to reveal their true chaotic nature later on.
Schwarzschild (1993) realized that chaotic orbits were all too frequent in triaxial
models and he tried to include them in his models but, then, the models did not
remain stationary. He computed models with orbits integrated over one Hubble time
and, then, new models using the same orbits but integrated between two and three
Hubble times. Differences beween equivalent models were revealed, e.g., by
differing axial ratios, as shown by Table~\ref{tab:dif}. 

\begin{table}[!ht]
\begin{center}
\caption{Axial ratio changes in the models of Schwarzschild (1993).}\label{tab:dif}
\begin{tabular}{@{}lccc@{}}
\hline
\hline
\rule{0pt}{1.05em}%
   $c/a$   & $b/a$ &  $\Delta (c/a)$ & $\Delta (b/a)$\\
 \hline
\rule{0pt}{1.05em}%
 0.7 &  0.8631 & -4.3\% & -7.3\%\\
 0.5 & 0.7906 & 2.6\% & -3.9\%\\
 0.3 & 0.7382 & 10.0\%& 0.9\%\\
 0.3 & 0.9534 & 4.3\% & 0.6\%\\
 0.3 & 0.4254& 16.7\% & 16.4\%\\
 0.3 & 0.3289& 6.7\% & 3.4\%\\
\hline
\end{tabular}
\end{center}
\end{table}

Of course, it would be obviously hopeless to try to use observations to detect such
small changes, so that we might ask: If we cannot recognize a truly stable galaxy
from a slowly evolving one, why care? Nevertheless, from a theoretical point of view,
it is certainly important. Therefore, the question we should be actually asking is: Is
it possible to have a stable self-consistent system if a significant fraction of its
material is in chaotic motion? No doubt, most dynamicists will feel inclined to
answer "No" to that question (see, e.g., Siopis and Kandrup 2000) but, as I will show,
they will be wrong.

From the observational point of view there is proof, both statitical (e.g., Ryden
1996) and on individual galaxies (e.g., Statler {\it et al.}), that at least some
elliptical galaxies are triaxial, and not merely rotationally symmetric (either
oblate or prolate). Moreover, the surface brightness of elliptical galaxies
increases towards the center forming a "cusp" (e.g., Crane et al. 1993, Moller et al.
1995) that reveals the presence of central mass concentration and, probably, a
black hole. The problem is that it has long been known (see, e.g., Gerhard \& Binney
1985) that central cusps and black holes can disrupt the box orbits that are necessary
to keep the triaxial form. The strong central field reorients the box and chaotic
orbits that come close to the center, so that a triaxial model will quickly evolve
into a rotationally symmetric one (we will return to this point later on). Is it
then possible to have cuspy triaxial galaxies? Can such galaxies have significant
amounts of matter on chaotic orbits? These questions are obviously important from
an observational point of view and for fitting models to the observations.

\subsection{The problem of diffusion}
Several authors haven been worried by the "problem" of chaotic mixing and, for that
reason, they have tried to avoid including chaotic orbits in their models. For
example, Merritt and Fridman (1996) refer to solutions that contain chaotic orbits
as {\it quasi equilibrium} and indicate that in quasi equilibrium models chaotic
mixing will produce a slow evolution of the model figure, specially near the center.
That idea comes from the assumption that, in the long run, chaotic orbits will
cover uniformly all the space available to them by the energy integral, i.e., a
region much rounder than the triaxial model, but that assumption is simply wrong.
Chaotic orbits do not occupy all the space available to them and live there
forever happily. First, they do not occupy all the space allowed to them by the energy
integral because there are usually "islands of stability" where chaotic orbits cannot
enter and, second, they may at times behave similarly to regular orbits. The more or
less chaotic behavior of a chaotic orbit can be easily followed, for example, computing
the finite time Lyapunov numbers (FTLNs hereafter) that increase, or decrease, as the orbit
behaves more, or less, chaotically. As shown by Muzzio {\it et al.} (2005), if one sorts
the fully chaotic orbits according to the values of those numbers, one finds that the
orbits with lower values give a more triaxial distribution than the orbits with higher
values.

The actual problem posed by the changing behavior of chaotic orbits is the one they
pose to the method of Schwarzschild. The method will favor the inclusion of orbits
elongated in the direction of the major axis but, when those orbits begin to behave
more chaotically, the shape of the model will become rounder. This effect could be
compensated if the model includes chaotic orbits that initially had a rounder
distribution and, later on, a more elongated one. In other words, one might have a
{\it dynamic equilibrium}: a given chaotic orbit will not occupy always the same
region of space (as regular orbits do) but, as it moves out to occupy a different zone,
another chaotic orbit will fill in the region left vacant by the former. I will show
later on that it is perfectly possible to achieve such dynamical equilibrium and to
get stable models that include large fractions of chaotic orbits, but one has to resort
to methods different from Schwarzschild's. One cannot see any simple way to take
into account the abovementioned effect in that method, and the situation is even worse
when constant axial ratios are used throughout the model. As indicated by Muzzio 
{\it et al.} (2005), in that case one has no rounder halo that could act as a reservoir
of chaotic orbits that, as time goes by, become more elongated and compensate for
those that, on the contrary, become rounder. In fact, stable triaxial models that
include high fractions of chaotic orbits become rounder as one goes farther from the
center of the system (see, e.g., Muzzio {\it et al.} 2005, Aquilano {\it et al.} 2007).

\subsection{Central black holes}
The effects of a central growing black hole on the dynamics of a triaxial system were
well described by Merritt and Quinlan (1998), who built an $N$-body model of a triaxial
galaxy and verified that it was stable. They then let a black hole grow at the center
of the model and, as could be expected, the central density distribution became much
steeper. But it also turned out that the model lost its triaxiality and the evolution
was faster for more massive black holes. The central regions became rounder, but even
the outermost regions were affected.

Rather different results were obtained by Poon \& Merrit (2002, 2004) who obtained
equilibrium models of the central regions of triaxial galaxies containing black holes
and checked their stability. They were able to find stable configurations that persisted
even within the sphere of influence of the black hole. Interestingly, their models
included chaotic orbits as well as regular ones. These results seem to be at odds
with other work, like that of Merritt \& Quinlan (1998). It is possible that the
fact that the models were built including the black hole, rather than letting it
grow after the model was built, might help to explain the difference.

\section{Chaotic orbits in elliptical galaxies}
The presence of chaotic orbits in triaxial potentials was noticed early on, even in the
original paper of Schwarzschild (1979). Moreover, in Schwarzschild (1993), which dealt with
the singular (i.e., cuspy) logarithmic potential, he showed that many orbits that resulted
from his "initial condition spaces" were indeed chaotic.

More recently, after the acceptance that elliptical galaxies are cuspy and that most of them
may be harboring central black holes, it was found that both cuspiness and central black
holes enhance the chaotic effects. These studies of orbits are usually performed on fixed
and smooth potentials, such as the one arising from the triaxial generalization of the model
of Dehnen (1993), whose density distribution is:

\medskip
\begin{equation}
\label{(Ec. Dehnen2)}
         \rho(m) =\frac{(3 - \gamma)aM}{4\pi} m^{- \gamma}(m + d)^{-(4-\gamma)}  
         \;  \; \;  \;{\rm with}\;  \; \;  \;  0 \leq \gamma < 3,
\end{equation}
\medskip

where $m$ is the triaxial radius already defined by equation (2), $M$ is the total mass,
$d$ is a scale parameter proportional to the effective  radius (i.e., the radius containing
half of the mass) and $\gamma$ parametrizes the slope of the central cusp. Merritt and
Fridman (1996) provided the corresponding equations for the potential, the forces and the
derivatives of the forces (which are needed for the variational equations that allow the
computation of the Lyapunov exponents). Those expresions involve rather complicated integrals
and must be solved numerically.

The axial ratios $b/a$ and $c/a$ give an idea of the flatness of the model, and it is usual
to measure the triaxiality, $T$, as:

\medskip
\begin{equation}
\label{(Ec. Triaxiality)}
         T =\frac{(a^2 - b^2)}{(a^2 - c^2)}.   
\end{equation}
\medskip

It goes from 0, for an oblate spheroid, to 1, for a prolate spheroid, and the ellipsoids with
$T = 0.5$ are called "maximally triaxial" ellipsoids.

Merritt and Fridman (1996) investigated maximally triaxial models with $\gamma =1$ (weak-cusp)
and $\gamma = 2$ (strong-cusp). Their libraries of orbits contained high fractions of chaotic
orbits, particularly in the strong-cusp case and for orbits with zero initial velocity. Thus,
it is not surprising that the attempts to build stable models without chaotic orbits were
doomed to failure. Using the method of Schwarzschild, Merritt and Fridman managed to build
"fully mixed" models for their weak-cusp case, but not for the strong-cusp one.

\subsection{The contribution of Kandrup and Siopis}
Siopis and Kandrup (2000) and Kandrup and Siopis (2003) did two very comprehensive investigations
of orbits in the triaxial Dehnen potential, allowing also for the presence of a central black
hole. They considered different axial ratios, cusp slopes and black hole masses and, in addition
to the chaoticity of the orbits, they also investigated chaotic diffusion and how it is affected
by noise.

They found that chaotic orbits tend to be extremely sticky for all cusp slopes, but specially
for the steepest ones. This fact was revealed by visual inspection of the orbits, but also by
the bimodal distribution of the FTLNs. Besides, if the orbits were not sticky, the FTLNs should
fall as $t^{-1/2}$, where $t$ is the integration time, but they found a much lower slope,
corroborating the stickiness. Moreover, the presence of a black hole increased the stickiness
of the orbits. Except for the steepest cusp, the $\it fractions$ of chaotic orbits were not
much affected by either the steepness of the cusp or the mass of the black hole. Nevertheless,
the $\it values$ of the FTLNs were significantly affected by both: steeper cusps and more massive
black holes led to larger FTLN values. They investigated the effect of triaxiality adopting
$a = 1.0$, $c = 0.5$ and selecting different $b$ values. As could be expected, the fractions of
chaotic orbits decreased both toward $b = 0.5$ (prolate spheroid) and toward $b = 1.0$
(oblate spheroid). Besides, going toward the innermost regions of the model the fraction of
chaotic orbits increased, and also increased the effect of the mass of the black hole on that
fraction. They also found that prolate spheroids tend to have larger fractions of chaotic
orbits, and with larger FTLNs, than oblate spheroids; that effect was more pronounced for the
outermost shells. Finally they considered changing both $b$ and $c$ adopting $b = 1 - D$ and
$c = 1 - 2D$, i.e., from a spherical system for $D = 0$, to a disk for $D = 0.5$. The fraction
of chaotic orbits turned out to increase monotonically with $D$, and even a very small departure
from sphericity yielded chaos. Without a black hole, the size of the FTLNs also increased
monotonically with $D$. Larger black hole masses resulted in larger FTLN values, but the increase
with $D$ leveled off for the most massive ones.

Chaotic mixing is another interesting aspect of the work of Kandrup and Siopis. They showed that
different ensembles of chaotic orbits diffuse in different ways and end up occupying different
regions even after long times. A very interesting aspect of their work is that they recognized
that galaxies are subject to several perturbations that can be modelled as "noise" and profoundly
affect chaotic diffusion. They considered: a) Periodic driving, to simulate the effect of a
satellite orbiting the galaxy; b) Friction and white noise to simulate discreteness effects;
c) Coloured noise to simulate the effects of encounters with other galaxies. They showed that all
these effects tend to increase the diffusion rate.

\subsection{Partially and fully chaotic orbits}
Since we are interested in equilibrium models of elliptical galaxies, the energy integral always
holds. Regular orbits have two additional isolating integrals, but we can have chaotic orbits
either without any other isolating integral (fully chaotic orbits), or with just one (partially
chaotic orbits). This difference had been recognized long ago (Contopoulos {\it et al.} 1978, Pettini
\& Vulpiani 1984), but little importance was assigned to it in galactic dynamics studies despite being
found in several investigations of triaxial systems (e.g., Goodman and Schwarzschild 1981, Merritt
and Valluri 1996). These studies used a sort of three dimensional (3-D) Poincar\'e map. If the velocity
on an orbit is computed every time that the particle returns to the same point, then: a) The velocities
of fully chaotic orbits will adopt any possible direction; b) Those of partially chaotic orbits
will fall on a curve; c) The velocities of regular orbits will have always the same, or at most a
finite number, of directions. Needless to say, this sort of 3-D Poincar\'e map is extremely computer
time consuming, because one has to follow the orbit over very long time intervals to have a significant
number of returns (almost) to the same point. A more efficient, albeit also computer time consuming,
method is to compute all (and not just the largest) Lyapunov exponents: the exponents of regular
orbits are all zero, partially chaotic orbits have one positive exponent, and fully chaotic orbits
have two positive exponents. Using Lyapunov exponents Muzzio (2003) showed that, in triaxial models
of elliptical galaxies, the spatial distributions of partially and fully chaotic orbits are diferent,
so that it is important to separate them in studies of galactic dynamics. Similar results were
obtained by Muzzio and Mosquera (2004) with models of galactic satellites and by Muzzio {\it et al.}
(2005), Aquilano {\it et al.} (2007), Muzzio {\it et al.} (2009) and Zorzi (2011) with models of
elliptical galaxies.

The Lyapunov exponents should be obtained integrating the orbit over an infinite time interval but,
as that is impossible for numerically integrated orbits, the FTLNs are used instead. That rises the
problem that one cannot reach zero values, even for regular orbits: if $T$ is the integration interval,
then the lowest FTLN values will be of the order of $\ln T/T$. Besides, one might ask which is the
practical limiting value to separate regular from chaotic orbits. The inverse of the Lyapunov exponent
is called the Lyapunov time, and it gives the time scale for the exponential divergence of the orbit.
At first sight, orbits with Lyapunov times longer than the Hubble time could be regarded as regular,
but one should recall that what actually interests in galactic dynamics is the orbital distribution,
rather than the exponential divergence of orbits. Therefore, the right question to ask is which is the
limiting value of the Lyapunov exponents that separates orbits whose distribution is similar to that
of regular orbits from those that have a clearly different distribution.  Aquilano {\it et al.} (2007),
Muzzio {\it et al.} (2009) and Zorzi (2011) investigated this problem and found that, for their models,
the limiting value corresponds to a Lyapunov time about six or seven times larger than the Hubble time. In
other words, orbits not chaotic enough to experience significant exponential divergence on a Hubble
time are, nevertheless, chaotic enough to have a distribution significantly different from that of
regular orbits. It should be noted, however, that the computed fraction of chaotic orbits is not
substantially altered by the limiting value selected: for the models of the authors cited, adopting
the limit corresponding to six or seven Hubble times may result in a fraction of chaotic orbits of,
say, $40\%$ as compared to $35\%$ for the limit corresponding to one Hubble time.

\section{The $N$-body method}
The method of Schwarzschild (1979) is not the only one available to obtain equilibrium models of
elliptical galaxies. Actually, self-consistent models of triaxial systems had been obtained earlier
using the $N$-body method (see, e.g., Aarseth \& Binney 1978). One starts with a certain distribution
of mass points and follows its evolution using an $N$-body code to integrate the equations of motion.
A wise selection of the initial distribution, or a little tinkering with the code, allows one to
obtain a stable triaxial system whose self-consistency is assured by the use of the $N$-body code.

One can begin, for example, with a spherical distribution of mass points with very small velocities.
Gravity will force the collapse of such system and the radial orbit instability will lead it toward
a triaxial equilibrium distribution (see, e.g., Aguilar \& Merritt 1990). One can change the initial
velocity dispersion to obtain systems with different triaxiality: the smaller the dispersion, the
larger the triaxiality. This method has been used by Voglis {\it et al.} (2002), Kalapotharakos and
Voglis (2005), and by my coworkers and myself. Another possibility is to launch one stellar
system against another and let them merge, yielding a triaxial system, a strategy followed, e.g.,
by Jesseit {\it et al.} (2005). 

Code tinkering was used, for example, by Holley-Bockelmann {\it et al.} (2001) who started with a
Dehnen (1993) spherical distribution of mass points and added to the $N$-body code a fictious force
that slowly squeezed the system, first in the $z$ direction, and then in the $y$ direction. The
original $N$-body code was finally used to let the system relax toward a final triaxial equilibrium
distribution.

After building the triaxial system, one has to investigate its orbital structure. The $N$-body code is
no longer useful because its potential is neither smooth (because it is the sum of the potentials of
the individual particles) nor time independent (due to the statistical changes in the distribution
of the finite number of particles). Therefore, a smooth and time independent approximation should
be adopted for the potential and then, orbits in that potential can be computed using as initial
conditions the positions and velocities of all, or a random sample of, the mass points. One can then
separate regular from chaotic orbits, and partially from fully chaotic orbits, with an adequate
chaos indicator and, finally, the different kinds of regular orbits can be obtained with an automatic
classification code based, e.g., on the analysis of the orbital frequencies.

Both Schwarzschild's and the $N$-body method end up with the same result: an equilibrium self-consistent
triaxial system and the knowledge of its orbital content. The steps to reach those results are, however,
different. One uses the orbits to obtain the system in Schwarzschild's method, while in the $N$-body
method the system is built by the gravitational interactions mimicked by the $N$-body code and thereafter
one performs the orbital analysis.

\subsection{The work of the Athens' group}

Voglis {\it et al.} (2002) obtained two non-cuspy stable triaxial models, Q (Quiet) and C (Clumpy),
from cold collapses of initially spherical particle distributions. The fractions of chaotic orbits
were $32\%$ for the former and $26\%$ for the latter. A similar model was built and investigated by
Kalapotharakos and Voglis (2005), who found $32\%$ of chaotic orbits. An interesting point that they
raised is that the real particles in their model did not cover all the space that one could cover
with test particles. It is a natural consequence of self-consistency, that only allows the presence
of certain orbits, and Kalapotharakos and Voglis showed this effect beautifully on the frequency map.
A very clever aspect of their work is that they determined the orbital frequencies of the chaotic
orbits, in addition to those of the regular orbits, and showed that, due to the stickiness, many of
them fall close to the loci of the regular orbits on the frequency map. Again, their model is stable
but, after the sudden introduction in it of a central black hole, it quickly evolves toward an
oblate shape.

More recently, Kalapotharakos (2008) investigated the evolution of two stable, non-cuspy, triaxial
systems after the introduction of a central black hole. He linked that evolution to the presence of
chaotic orbits and introduced a parameter, dubbed effective chaotic momentum, that correlates well
with the rate of evolution of the system.

\subsection{The puzzles of the work of Holley-Bockelmann et al.}
As already mentioned, Holley-Bockelmann {\it et al.} (2001) obtained cuspy triaxial systems through
adiabatical deformation of Dehnen (1993) spherical models. Their systems turned out to be very
stable, but they found very little chaos (less than $1\%$, and they even attributed that little
to noise in their potential approximation), in complete contradiction with the results of the
Athen's group and our own. Kandrup \& Siopis (2003) pointed out that the Fourier technique used
by Holley-Bockelmann {\it et al.} could have made them miss many chaotic orbits, a likely possibility
considering that such technique is not as good as others for chaos detection (see,
e.g., Fig. 2 of Kalapotharakos and Voglis 2005). Nevertheless, one should recall that the collapses
used by the Athens' group and ourselves to build triaxial models yield predominantly radial orbits,
which have long been known to favor chaos (see, e.g., Martinet 1974), while the adiabatical squeezing
used by Holley-Bockelman {\it et al.} probably resulted in a more isotropic velocity distribution,
so that the discrepancy among the chaotic fractions might be real.

In a second paper, Holley-Bockelmann {\it et al.} (2002) let a black hole grew at the center of
one of their models. As could be expected, they found that the central cusp became steeper and
the central regions rounder. But the outer regions (even those that contained the inner $10\%$
of the total mass) retained the triaxial shape. Just as mentioned before about the work of
Poon and Merritt (2002, 2004), these results seem to be at odds with other investigations, like
that of Merritt \& Quinlan (1998).

\subsection{The work of our La Plata-Rosario group}
In our first works we used 100,000 particles to build our $N$-body models, and for the past few
years we have been using one million, guaranteeing in any case the accuracy of stability studies.
Besides, we typically classify between 3,000 and 5,000 orbits per model, which provides adequate
statistics. To build our models we just use the simple cold collapse technique and we do not try
to follow the actual galaxy formation process, because we are not interested in that process but
only in obtaining models morphologically similar to elliptical galaxies. For example, although it
is not physically realistic, we let our models relax during intervals of several Hubble times to
ensure that our subsequent studies refer to equilibrium models.

Muzzio {\it et al.} (2005) obtained a model of a non-cuspy E6 galaxy that contained $52.7\%$ of
chaotic orbits. Later on, Muzzio (2006) found that the model displayed very slow figure rotation
(i.e., it rotated despite having zero angular momentum). Taking the rotation into account, the
fraction of chaotic orbits raised to $56.6\%$, probably because of the symmetry breaking caused
by the rotation, despite its small value. Later on, Aquilano {\it et al.} (2007) obtained three
models of non-cuspy galaxies resembling elliptical galaxies of types E4 through E6, all of them
very stable and with between one third and two thirds of chaotic orbits. All these models
had axial ratios that increased from the center toward the outer parts of the galaxy, which
may have contributed to them being stable and chaotic at the same time. Besides, all of them
corroborated the need to distinguish partially from fully chaotic orbits because their spatial
distributions were different.

Muzzio {\it et al.} (2009) managed to obtain cuspy models of E4 and E6 galaxies that, again,
were highly stable despite having about two thirds of chaotic orbits. Nevertheless, those models had
pushed to the limit the use of the $N$-body code of Aguilar (see, White 1983 and Aguilar and Merritt
1990) and, in fact, they had to introduce a small additional potential to compensate for the softening
needed by the code. Therefore, in order to continue investigating cuspy models, we switched to the
code of Hernquist (Hernquist and Ostriker 1992) that needs no softening and uses a radial expansion
of the potential in functions related to the potential of Hernquist (1990) which is itself
cuspy with $\gamma = 1$. Our first work with this code is the PhD thesis of Zorzi (2011), defended
last June at the Universidad Nacional de Rosario.

Kalapotharakos {\it et al.} (2008) investigated the approximation of $N$-body realizations of
models of Dehnen (1993) for different values of $\gamma$ using the self-consistent field method
(the method of Hernquist \& Ostriker 1992 is just a special case of that method), and they found
that the choice of the radial basis functions seriously affected the results obtained on
chaotic orbits. Nevertheless, the method of Hernquist \& Ostriker turned out to be adequate
for models with $\gamma \simeq 1$, and we had selected the number of terms in the potential
expansion so as to have models whose cusps had that slope (Zorzi \& Muzzio 2009).

\begin{figure}[!ht]
  \centering
  \includegraphics[width=.75\textwidth]{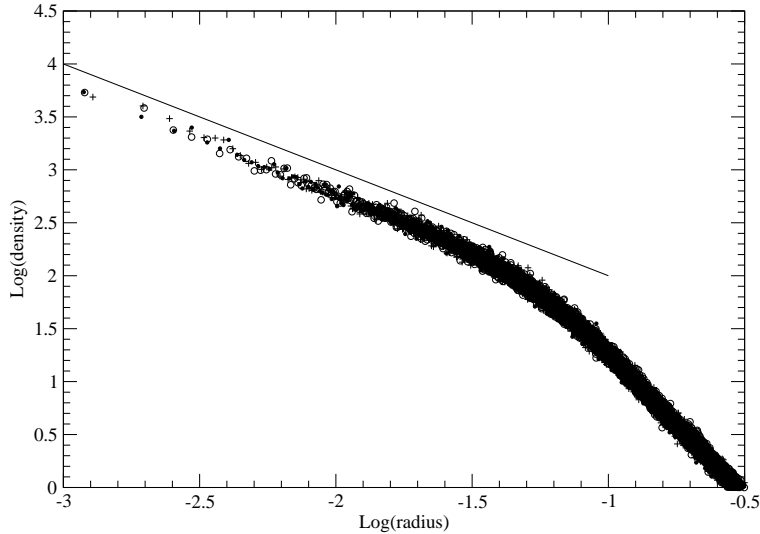}
  
  \caption{The logarithm of the density vs. the logarithm of the radius for the E4 models
of Zorzi (2011). Filled and open circles and crosses correspond to three statistically equivalent
realizations of the same model. The slope of the straight line is equal to 1.}
  \label{fig:ab1}
\end{figure}

Zorzi (2011) built four groups of models that mimic E2, E3, E4 and E5 galaxies. Each group
consisted of three models that differed only in the seed number used to randomly generate
the initial conditions. In other words, for each galaxy type she had three statistically
equivalent models, which is important to control the consistency and robusteness of the
results to be obtained from them. In fact, except for the figure velocity rotation, all
her results were essentially the same for statistically equivalent models. Figure~\ref{fig:ab1}
gives the logarithmic density vs. radius plot for the central regions of her E4 models
and we can notice the excellent agreement among the different realizations of the model,
as well as the $\gamma \simeq 1$ cusp.

To check the stability of the models, she computed their central density and their three moments
of inertia over intervals of the order of five Hubble times. The changes in those quantities
were very small, indeed, smaller than $2.7\%$ in one Hubble time in all cases. Moreover, she
showed that most of that change is most likely due to relaxation effects in the $N$-body code
(see Hernquist \& Barnes 1990). Not only are these values smaller than most of those of
Schwarzschild shown in Table~\ref{tab:dif}, but Zorzi's had been obtained with self-consistent
models, while Schwarzschild's correspond to a fixed potential. Actually, when the potential is
fixed, the changes in the models of Zorzi are an order of magnitude smaller than those obtained
self-consistently. The fractions of chaotic orbits are extremely high in Zorzi's models, up to
more than $85\%$ for her E2 and E5 models, and larger that $75\%$ for the other two. The values
of the FTLNs are also very high, almost twice the values of the equivalent non-cuspy models of
Aquilano {\it et al.} (2007). And, as in our previous work, she also found that partially and fully
chaotic orbits have different distributions and should be analyzed separately.

\section{Conclusions}
The works of the Athens' group and our own show that it is perfectly possible to build triaxial
stellar systems, even cuspy ones, that include large fractions of chaotic orbits and are,
nevertheless, highly stable over intervals of the order of one Hubble time. All these models
were obtained using the $N$-body method, so that the difficulty to obtain such models with the
method of Schwarzschild should be attributed to the method itself and not to physical causes.

We certainly need more models built with the adiabatic squeezing technique of Holley-Bockelmann
{\it et al.} (2001). The orbits in those models are probably less radial than those in the
models arising from cold collapses and that might result in lower fractions of chaotic orbits
and models more capable of preserving triaxiality when harboring a black hole.

The distributions of partially and fully chaotic orbits are certainly different, but much remains
to be done on this subject. Are the partially chaotic orbits merely confined to the stochastic
layers around resonances? Or do they fill in connected regions where an isolating integral, or
pseudo integral, holds? And, certainly, we need better and faster methods than that of the
Lyapunov exponents to separate partially from fully chaotic orbits.

Modern observations suggest not only that elliptical galaxies have significant rotation, but that
they even rotate in different directions at different distances from the center, perhaps a consequence
of past mergers. Nevertheless, there are not many works on rotating ellipticals (a recent
exception is that of Deibel et al. 2011) and more work is clearly warranted on this subject.

The figure rotation we found in many of our models is certainly puzzling. All the tests we made
support that it is real and, besides, these models can be seen as the stellar counterpart of the
Riemann ellipsoids of fluid dynamics. But why, appart from the tendency of flatter systems to
rotate faster, the rotation shows no obvious connection to the properties of the model and, worst,
why do statistically equivalent models display different rotation velocities? This is another
subject where more research is clearly warranted.

Last but no least, let me end recalling that, although I have only briefly mentioned regular
orbits here, their study is also very interesting. Not only are they important for the dynamics
of elliptical galaxies, but they also offer technical challenges, like the design of fast and
accurate methods of automatic classification.

\acknowledgements I am very grateful to C. Efthymiopoulos for useful comments and to
R.E. Mart\'{\i}nez and H.R Viturro for their technical assistance. This work was
supported with grants from the Consejo Nacional de Investigaciones Cient\'{\i}ficas
y T\'ecnicas de la Rep\'ublica Argentina, the Agencia Nacional de Promoci\'on
Cient\'{\i}fica y Tecnol\'ogica and the Universidad Nacional de La Plata.

\end{document}